\documentclass[a4paper,10pt]{article}

\title{Toward reconstruction of relative state formulation of quantum theory}

\author{Yukinari Kurita\footnote{e-mail:ykurita@phys.ualberta.ca} \\ Department of Physics, P-412, Avadh Bhatia Physics Laboratory,\\ University of Alberta, Edmonton, Alberta, Canada, T6G 2J1 }
\date{ April 5, 2004 }

\begin{document}
\maketitle
\begin{abstract}
In quantum theory, it is widely accepted that all experimental results must agree with theoretical predictions based on the Copenhagen interpretation. However the classical system in the Copenhagen interpretation has not been defined yet. On the other hand, although ongoing research of decoherence is trying to elucidate the emergence of the classical world, it cannot answer why we observe one of eigenstates in observed system. These situations show that the relation between what we observe and physical law has not been elucidated. Here I elucidate the relation by developing Everett's suggestion. Further, from this point of view, I point out that today's brain science falls into circular argument because it is trying to assign what we observe in the brain to process of the subjective perception, and I suggest the future research line in brain science.
\end{abstract}

\section{Introduction}

In quantum theory, it is widely accepted that all experimental results must agree with theoretical predictions based on the Copenhagen interpretation. However the classical system in the Copenhagen interpretation has not been defined yet. On the other hand, ongoing research of decoherence is trying to elucidate the emergence of the classical world. However the research of decoherence can not answer why we observe one of eigenstates in observed system. These situations show that the relation between what we observe and physical law has not been elucidate yet. 

On the other hand, today's brain science is trying to assign what we observe in the brain to process of the subjective perception. However, as I mentioned, the relation between what we observe and physical law has not been elucidated. Therefore today's brain science is not based on physical law, but is based on information processing of the brain which has not been elucidated yet. Therefore today's brain science falls into circular argument. 

Here I elucidate the relation between what we observe and physical law by developing Everett's suggestion or relative state formulation of quantum theory. This is an attempt to reconstruct relative state formulation of quantum theory. From this point of view, I point out what today's brain science misses, and I suggest the future research line in brain science. 

\section{Observation problem of quantum theory}

The formulation of observation of quantum theory is the following:

A physical system is described by a state vector $\mid\psi\rangle$,which is an element of a Hilbert space. There are two fundamentally different ways in which the state vector can change.

\begin{description}
\item[Process 1:] The discontinuous change brought about by the observation of a quantity with eigenstates $\mid\phi_1\rangle$,$\mid\phi_2\rangle$,\ldots,in which the state $\mid\psi\rangle$ will be changed to the eigenstate $\mid\phi_i\rangle$ with probability $\mid\langle \phi_i\mid\psi\rangle\mid^2$ .

\item[Process 2:] The continuous, deterministic change of state of an isolated system ,which is described by a state vector $\mid\psi\rangle$ ,with time according to the following equation: $i\frac{\partial \mid\psi\rangle}{\partial t}=H\mid\psi\rangle$
\end{description}

It has been pointed out that the discontinuous change described by Process 1 can't be derived by the continuous change described by Process 2. That is, if the change with time of the state of the system which consists of not only the observed system described by state vector$\mid\psi\rangle$ but also observing system(apparatus, observer) is described by the above Process 2, it is impossible to describe the change with time of the state of the observed system by the above Process 1. This is what is called observation problem of quantum theory.

Apart from the above problem, however, it is widely accepted that all experimental results must agree with theoretical predictions based on the Copenhagen interpretation. According to this interpretation, the change with time of a quantum system is described by the above Process 2, and the change described by the above Process 1 is caused by the interaction between the quantum system and a classical apparatus.

However, the Copenhagen interpretation can't answer what the classical system is. There is no universally accepted definition of the borderline between the quantum system and the classical system \cite{zurek91, braginsky87}.

Not only can't Quantum theory (the Copenhagen interpretation) answer what the classical system is, it also can't answer what the observed system is. In fact, although quantum theory defines the mechanical variable that can be observed as observable, it can't answer what defines the quantum system whose change is described by the above Process 1.

To be more precise, the observed system is merely defined intuitively in real quantum measurement. A measurement is a series of many physical processes which subsequently take place in the observed system, apparatus, and observer. Among the series of many physical processes we can almost always find the process which can be described by the above Process 1. In other words, there is the following rule for placing the above Process 1, "put sufficiently much into the quantum system that the inclusion of more would not significantly alter practical prediction. \cite{braginsky87,bell87, zeh73}

From the above paragraph, it can be concluded that although, in real quantum measurement, there is the quantum system whose change is described by the above Process 1 accurately enough, quantum theory can't answer how the system is defined . 

\section{Relative state formulation}

Once, Everett \cite{everett57} suggested that there is only the change described by Process 2 in the physical world, and that we do not need Process 1 when we describe an observation process. Further, he insisted that the observed system after observation exists merely as the state relative to an observer(Relative state). That is, if the observer described by the state $\mid O\rangle$ observes one of eigenstates [$\mid s\rangle$] of state $\mid\psi\rangle$, after the interaction described by the following equation, the state of the observer who observed certain eigenstate $\mid s\rangle$ is described by $\mid O_s\rangle$. Further, in this observer's world, there is only the Relative state $\mid s\rangle$.

\begin{equation}
\label{myeqn1}
\mid\psi\rangle\bigotimes \mid O\rangle=[\sum_{s}c_s\mid s\rangle]\bigotimes \mid O\rangle\\
\to\sum_{s}c_s\mid s\rangle\bigotimes \mid O_s\rangle
\end{equation}

In the Everett's paper, we never find the comment ``the splitting of the world''. It is the observer that splits. That is, in the case of the above equation (\ref{myeqn1}), the state of the observer $\mid O\rangle$ splits into each observer's state $\mid O_i\rangle$,$\mid O_j\rangle$, \ldots. In other words, the splitting described by the following equation (\ref{myeqn2})  never happens. Here, $\mid R\rangle$ is the rest of the world state, which doesn't contain the state of the observer $\mid O\rangle$ and the state of the observed system $\mid\psi\rangle$.

\begin{equation}
\label{myeqn2}
\sum_{s}c_s\mid s\rangle\bigotimes \mid O_s\rangle\bigotimes \mid R_s\rangle
\end{equation}

However, there are some problems. Now I explain this by an model that an observer observes an observed system; Here both the observer and the observed system are assumed to be stable matters. Further, the observer is assumed to be described by a macroscopic localized state $\mid \Psi \rangle$, and the observed system is assumed to be described by superposed states of macroscopic localized state $\mid \psi_- \rangle $ and $\mid \psi_+ \rangle $, which are spatially separated each other. As the observation process, the observer is assumed to interact with the observed system through photons. To be more precise, photons are scattered by the observed system, and then the scattered photons interact with the observer.

At first step, let me think that the observer interacts with an observed system whose state is $\mid \psi_- \rangle $. After interaction, states of the observed system, observer, and photons are not separable. So it is difficult to define each system \cite{rossler96}. However, if both the observer and the observed system are stable, the change of states of the observer and the observed system is negligible. Therefore the interaction process could be described by the following. Here observer's state $\mid \Psi_- \rangle $ correlates with the observed system, and its change from the state $\mid \Psi \rangle $ is very small.
\begin{eqnarray}
\label{myeqn3}
\mid \psi_- \rangle \mid \Psi \rangle \to \mid \psi_- \rangle \mid \Psi_- \rangle 
\end{eqnarray}

Similarly, the process which the observer interacts with an observed system whose state is $\mid \psi_+ \rangle $ could be described by the following. Here observer's state $\mid \Psi_+ \rangle $ correlates with the observed system, and its change from the state $\mid \Psi \rangle $ is very small.

\begin{eqnarray}
\label{myeqn4}
\mid \psi_+  \rangle \mid \Psi \rangle \to \mid \psi_+ \rangle \mid \Psi_+ \rangle 
\end{eqnarray}

From the above two processes, we can guess the process that the observer interacts superposed states of the observed system. Roughly speaking, it could be expressed by adding the above two equations(\ref{myeqn3}, \ref{myeqn4}). ( Here, I neglect the interaction between state $\mid \psi_- \rangle $ and state $\mid \psi_+ \rangle $.)

\begin{eqnarray}
\label{myeqn5}
(\mid \psi_- \rangle +\mid \psi_+ \rangle) \mid \Psi \rangle \to \mid \psi_- \rangle \mid \Psi_- \rangle +\mid \psi_+ \rangle \mid \Psi_+ \rangle 
\end{eqnarray}

In the above equation, state of the observed system $\mid \psi_- \rangle $ and $\mid \psi_+ \rangle$ is defined as relative state of the observer's state $\mid \Psi_- \rangle $ and $\mid \Psi_+ \rangle $, respectively. However this is not the whole story \cite{zurek81}. We can expand the state of the observed system ($\mid \psi_- \rangle +\mid \psi_+ \rangle $) into the following way. Here $\mid w_n \rangle $ and $a_n $ are energy eigenstate of the observed system and its weight, respectively.

\begin{eqnarray}
\label{myeqn6}
\mid \psi_- \rangle +\mid \psi_+ \rangle = \sum a_n \mid w_n\rangle 
\end{eqnarray}

In this case, the above process can be described by the following. Here the observer's state $\mid \Psi_n \rangle$ correlates with the state $ \sum_{i} b_{in} \mid w_{i} \rangle $, and its change from the state $\mid \Psi \rangle $ is very small.
\begin{eqnarray}
\sum a_n \mid w_n\rangle \mid \Psi \rangle \to \sum_{i,n} a_n b_{in} \mid w_{i} \rangle \mid \Psi_n \rangle \\
\mathrm{provided;}\ \mid w_n \rangle \mid \Psi \rangle \to \sum_{i} b_{in} \mid w_{i} \rangle \mid \Psi_n \rangle 
\end{eqnarray}

In the above equation, state of the observed system is defined as relative state of the observer's state $\mid \Psi_n \rangle $. Therefore, it is not clear what state the observer observed. 

However, the above discussion misses decoherence of the observed system. Macroscopic superposition of different spatially localized states decoheres by coupling with its environment \cite{joos96, joos85}. Therefore the expansion (\ref{myeqn6}) is impossible. Assuming that the environment is photons which were scattered by the observed system long time ago, the interaction process is thought to be similar to the process of the equation (\ref{myeqn5}) even though the composed system of the observer, the observed system and its environment is not isolated system. So the interaction between the observer and the observed system is the following. 

\begin{eqnarray}
(\mid \psi_- \rangle \mid \Xi_- \rangle +\mid \psi_+ \rangle \mid \Xi_+ \rangle) \mid \Psi \rangle \\
\to \mid \psi_- \rangle \mid \Xi_- \rangle \mid \Psi_- \rangle +\mid \psi_+ \rangle \mid \Xi_+ \rangle \mid \Psi_+ \rangle 
\end{eqnarray}

Before the interaction, state of the environment $\mid \Xi_- \rangle $ and $\mid \Xi_+ \rangle $ correlates with state of the observed system $\mid \psi_- \rangle $ and $\mid \psi_+ \rangle $, respectively. After the interaction, state of the observed system is well defined as relative state of the observer's state. In other words, the observer splits when the observer interacts with the observed system whose superposed states were decohered by coupling with its environment 

However, it is not the only condition that the observer splits. I show another example; The process that the observer scatters photon is described in the following. Here $ h_n(\vec{x})$, $c_n$, $ \mid n \rangle$, $\mid j \rangle$ and $ S_{lm,nj}$ are momentum eigenfunction of the observer, its weight, momentum eigenstate of the observer, state of the photon and scattering matrix.

\begin{eqnarray}
\mid \Psi \rangle \mid j \rangle \equiv \sum_{n}c_nh_n(\vec{x}) \mid n \rangle \mid j \rangle\\
\to \sum_{nlm} c_nh_n(\vec{x}) S_{lm,nj}\mid l \rangle \mid m \rangle 
\end{eqnarray}

In the above process, if the observer is stable, change from the state$\mid \Psi \rangle$ to the state $ \sum_{nlm} c_nh_n(\vec{x}) S_{lm,nj}\mid l \rangle $ is very small. However after the scattering, if each state $\mid m \rangle$ of the scattered photon decoheres, the observer splits into the system each of which correlates with each decohered state of photon.

From this point of view, the following is concluded; 

\begin{description}
\item[Conclusion 1. Split;] the observer splits whenever the observer interacts with an system whose superposed states were decohered by coupling with its environment, and whenever superposed states of the system which interacted with the observer decohere. 

\item[Conclusion 2. Observation;] observation is selective interaction between an observer and an observed system which is cut from its environment by certain physical process.
\end{description}
In principle, the observer is any stable matter, whose change by the interaction with its environment is negligible. However there are 2 problems. First of all, there is no rules which distinct certain system from its environment in quantum theory. Secondly, as I pointed out in the above, what defines quantum system whose change is described by the above Process 1 has not been solved yet. The solution of the first problem solves second problem.

We need observer who has an ability to distinguish itself from its environment. It is something new for quantum theory. It is consciousness that has this ability. Although it seems to contradict our intuition, consciousness has a role to define an observer which cannot be done by quantum theory. It is not until consciousness defines an observer that observed system and its environment are defined because quantum theory cannot define them.

Therefore, it is concluded that there is a stable matter which distinguishes itself from its environment in the brain. Here, I define that matter as MSC(material substrate for consciousness); That an observer perceives an observed system is the phenomena that the MSC in the brain interacts with the observed system and the MSC correlates the observed system.

Through the MSC, we can solve the second problem. It has been pointed out that the Process 1 could occur much later in the observational chain than decoherence \cite{zeh55}. Does this statement contradict the above Conclusion 1? Answer is no. The reason is that the human's ability to cut the observed system from its environment is not so fine. We observe outside world by using visible ray, so the ability to cut the observed system from its environment is limited.

From this point of view, the following is concluded.
\begin{description}
\item[Conclusion 3;] The classical system in Copenhagen interpretation is the quantum system observed by the MSC.
 
\item[Conclusion 4;] The reason that we don't observe superposed states of the macroscopic quantum system is that the MSC in our brain splits into MSCs each of which correlates each eigenstate of the system.
\end{description}

\section{Psychophysical parallelism}

Once von Neumann pointed out that the following principle of ``psychophysical parallelism'' is the fundamental requirement for the scientific viewpoint \cite{neumann55};
\begin{quote}
it must be possible so to describe extra-physical process of the subjective perception as if it were in reality in the physical world--i.e., to assign to its parts equivalent physical processes in the objective environment, in ordinary space.
\end{quote}

Today's brain science is based on the above mentioned ``psychophysical parallelism''. One of the prerequisites of applying this principal is to define ``physical processes in the objective environment''. However, there has not been a rigorously scientific definition on this concept, and scientists have used this concept as what they observe. However, it means that this concept is based on information processing of the brain, which has never been elucidated yet. Therefore today's brain science falls into a circular argument.

If we define ``physical processes in the objective environment'' as what we observe, we miss the important process of brain dynamics. In the brain, MSC must cut an system from its environment and split. However what we observe is the classical system which has been cut from its environment  by the observation process. To elucidate information processing of the brain, we must investigate what MSC is, and we must think the dynamics of the brain not from outside the brain but from view of MSC. 

As is often said, perception depends on the simultaneous, cooperative activity of neurons \cite{freeman91, edelman00}. This suggests that there are many MSCs in each brain. However, each MSC should be stable matter whose change by the interaction with its environment is negligible because it is one of the prerequisites of an observer. 

By the way, according to research of decoherence, delocalized coherence states of an quantum system is damped exponentially by interaction between the system and its environment \cite{joos96, tegmark93}. This means that coherence of superposed states of an system doesn't disappear strictly. Therefore, threshold of MSC's splitting must be defined even though it is negligibly small.

\section{Discussion}

One of the motivations of Everett's work was to apply quantum theory to the closed universe. However, is it possible to define state vector of the whole universe? In natural science, the researched system is usually regarded as independent existence and is usually described by external observers. It is doubtful whether the universe can be described by external observers. Does universality of quantum theory mean that the state of the universe can be defined? In the following, I suggest the outline of reasoning of the negative possibility.

The charge measurement of electron depends on the scale parameter, for example, the magnitude of the momentum of the photon used for the charge measurement. The scale parameter has no analogue in the classical electrodynamical Lagrangian. At this point, Shirikov \cite{shirkov88} pointed out the relation between the scale parameter and the following Bohr's idea; ``To specify the quantum system, it is necessary to fix its ``macroscopical surrounding'', i.e. to give the properties of macroscopical devices used in the measurement process.'' The scale parameter describes the macroscopic devices.There is the possibility that the bare electron is ether. Thinking of the bare electron is no better than thinking of the orbit of electron in an atom. For example, it might be impossible to define the bare electron which doesn't contain the self-field. The existence of an electron is observed as if the bare electron interacted with its self-field, and the self-field seems to exist in the energy range that is consistent with the observed energy of electron. 

To make relative state formulation take the place of Copenhagen interpretation, ``macroscopical devices'' should be defined, though macroscopical devices also contain electrons. It seems to be circular argument. To solve this problem, the following could be suggested; There is no independent system, and it is the interaction that guarantees mutual existence of the physical system. The physical system is defined only by defining mutual relation. From this point of view, to look for the theory which doesn't use the renormalization method is thought to be based on the implicit hypothesis that an elementary particle can be defined as an independent system. 

The above discussion suggests the possibility of the new way to research cosmology, which avoids describing the state of the universe. Further, from the above point of view, the relative state formulation could be reconstructed as completely internal observer's theory.

\subsubsection*{Acknowledgements}
The author would like to thank Don N.\ Page, Faqir Khanna, Jack A.\ Tuszynski, and Otto E.\ Rossler for useful comments. The author also would like to thank Department of Physics in University of Alberta for hospitality. 

\addcontentsline{toc}{section}{References}
\bibliographystyle{abbrv}

\end{document}